\documentclass[fleqn,usenatbib]{mnras}

\usepackage{newtxtext,newtxmath}
\usepackage{mathptmx}
\usepackage{hyperref}
\usepackage[T1]{fontenc}

\usepackage{graphicx}	
\usepackage{amsmath}	
\usepackage{amssymb}	

\title[Particle content and radio-galaxy morphology]{Particle content, radio-galaxy morphology and jet power: all
  radio-loud AGN are not equal} \author[J. H. Croston]
{J. H. Croston$^{1}$\thanks{Email: Judith.Croston@open.ac.uk}, J. Ineson$^{2}$,
and M.J. Hardcastle$^{3}$\\$^{1}$ School of Physical Sciences, The Open
  University, Walton Hall, Milton Keynes, MK7 6AA, UK\\$^2$ School of Physics and Astronomy,
University of Southampton, Highfield, Southampton SO17 1BJ, UK\\$^3$
Centre for Astrophysics Research, School of Physics, Astronomy and Mathematics, University of
Hertfordshire, College Lane, Hatfield, AL10 9AB, UK}
 
\pagerange{\pageref{firstpage}--\pageref{lastpage}}
\pubyear{2017}

\defcitealias{ineson17}{I17}
\defcitealias{ineson15}{I15}
\defcitealias{ineson13}{I13}
\hypersetup{draft}
\begin{document}

\maketitle

\label{firstpage}

\begin{abstract}
 Ongoing and future radio surveys aim to trace the evolution of black hole growth and feedback from active galactic nuclei (AGN) throughout cosmic time; however, there remain major uncertainties in translating radio luminosity functions into a reliable assessment of the energy input as a function of galaxy and/or dark matter halo mass. A crucial and long-standing problem is the composition of the radio-lobe plasma that traces AGN jet activity. In this paper, we carry out a systematic comparison of the plasma conditions in Fanaroff \& Riley class I and II radio galaxies to demonstrate conclusively that their internal composition is systematically different. This difference is best explained by the presence of an energetically dominant proton population in the FRI, but not the FRII radio galaxies. We show that, as expected from this systematic difference in particle content, radio morphology also affects the jet-power/radio-luminosity relationship, with FRII radio galaxies having a significantly lower ratio of jet power to radio luminosity than the FRI cluster radio sources used to derive jet-power scaling relations via X-ray cavity measurements. Finally we also demonstrate conclusively that lobe composition is unconnected to accretion mode (optical excitation class): the internal conditions of low- and high-excitation FRII radio lobes are indistinguishable. We conclude that inferences of population-wide AGN impact require careful assessment of the contribution of different jet sub-classes, particularly given the increased diversity of jet evolutionary states expected to be present in deep, low-frequency radio surveys such as the LOFAR Two-Metre Sky Survey.

\end{abstract}


\section{Introduction}
\label{intro}




Determining the particle content of radio galaxies has been a
long-standing challenge, as the radio synchrotron emission with which
they are primarily observed does not uniquely determine the source
internal energy. \citet{burbidge56} made the first estimates of the
energy content of a radio galaxy by assuming the minimum total energy that could produce the observed radio synchrotron emission. The minimum energy assumption is closely equivalent to assuming the equipartition of
energy between relativistic leptons and magnetic field, and such approaches have been widely used to estimate magnetic field strengths and radio source energetics. However, if radio-galaxy jets and lobes contain non-radiating particles (e.g. relativistic protons, or material entrained from the surrounding medium), then equipartition calculations may underestimate the total energy by a large factor. Even if non-radiating particles are accounted for, there is no firm theoretical basis for expecting equipartition to apply across the radio-galaxy population. 

With the current generation of X-ray observatories, {\it Chandra} and {\it XMM-Newton}, it has become possible to test the equipartition assumption rigorously, and to understand in what circumstances it applies. For the Fanaroff-Riley \citep{fanaroff74} II radio-galaxy population, it is now routinely possible
to detect X-ray inverse Compton emission from their radio lobes (this
is not the case for FRIs). Following a number of detections of individual sources or small samples 
\citep[e.g][]{isobe02,hardcastle02,comastri03,croston04}, \citet{croston05} and \citet{kataoka05} presented the first large samples of X-ray inverse Compton lobe
measurements, demonstrating that magnetic field strengths in FRII
radio galaxies are typically a factor of 2 - 3 below the equipartition
values (so that internal energies are typically a factor of $\sim 5$ above the
minimum energy). We have argued \citep[e.g.][]{croston05} that the
near equipartition $B$ fields for FRIIs would not be expected if they
were energetically dominated by protons. Further support
for this argument comes from our recent large environmental study of
the radio-galaxy population (\citealt{ineson15,ineson17}, hereafter \citetalias{ineson15} and \citetalias{ineson17}, respectively), which has
allowed us to carry out a well-constrained comparison of external and
internal pressures of the FRII population for the first time, using
X-ray inverse Compton measurements for the radio lobes to determine
internal energy, and a comparison of the internal pressure with the
external pressure from the surrounding group or cluster to determine
the lobe expansion speed. This work has firmly demonstrated that FRIIs are typically over-pressured at their tips relative to their external environment, without the need for non-radiating particles.

The situation is different for low-power (FRI) radio galaxies. Early X-ray measurements 
of external gas pressure acting on radio lobes \citep[e.g.][]{morganti88} revealed that 
FRI lobes at equipartition, with no non-radiating particles, would typically be under-pressured 
relative to the external medium. Subsequent work on small samples \citep[e.g.][]{worrall00,croston08} 
has confirmed that FRI radio galaxies cannot be correctly described by the minimum energy/equipartition 
condition in the absence of a substantial non-radiating particle population. A particularly extreme 
situation applies for some X-ray cavity systems in nearby galaxy clusters, where proton-to-electron energy ratios of $100-10^{5}$ appear to be required \citep{dunn04,birzan08}. While there are several possible explanations for the apparent departure from equipartition in FRI radio galaxies, we have shown that in the case of tailed FRIs the increasing pressure discrepancy along the tails is best explained by entrainment of material from the surrounding intragroup or intracluster medium \citep{croston14}.

X-ray studies therefore strongly suggest that the internal conditions of FRI and FRII radio galaxies are systematically different. To date, however, there has been no direct population-wide comparison of the energetics for the two classes of radio galaxy. Such a comparison is the purpose of this paper. We have constructed a large, representative sample of FRI and FRII radio galaxies with well-determined environmental gas distributions (\citetalias{ineson15}), and in this paper we compare the FRII results of \citetalias{ineson17} with
inferred pressure ratios of the FRI radio galaxies from the parent
sample of \citetalias{ineson15}, with the aims of: (1) establishing conclusively whether the particle content and energy division between
particles and magnetic field {\it must} be different for the FRI and
FRII radio-galaxy sub-populations, and (2) if so, what the
implications are for jet power estimation from radio surveys. In
Section~\ref{sec:pcont} we present new comparisons of pressure ratios
for FRI and FRII radio galaxies and discuss the implications for their
particle and energy content; in
Section~\ref{sec:qjet} we discuss the implications of our results for
jet-power scaling relations; and in Section~\ref{sec:accmode} we
briefly comment on whether our results indicate any relationship
between particle content and accretion mode.

\section{Using pressure ratios to diagnose lobe internal conditions}
\label{sec:pcont}
The ratio of internal radio-lobe pressure to external pressure from the environment is a useful measure of the lobe dynamics at the tip of the radio galaxy (where it is closely linked to the advance speed, as discussed in \citetalias{ineson17}). The pressure ratio at the lobe {\it midpoint} can also be used to investigate particle content: the ratio at the lobe midpoint should be at least unity for most sources, otherwise the lobe must be contracting -- this cannot be the case for the majority of sources. The aim of this work is therefore to use midpoint pressure ratios to compare systematically the internal to external pressure ratios for FRI and FRII radio galaxies. Our parent sample is that presented in the environmental study of \citetalias{ineson15}, and we adopt the FR classifications given by \citetalias{ineson15} and \citetalias{ineson13}; however, we have excluded 3C\,305, which was included in the \citetalias{ineson15} study, but has a peculiar morphology intermediate between FRI and FRII \citep{hardcastle12}. 

\subsection{A comparison of FRI and FRII pressure ratios}
\label{sec:comp}
In \citetalias{ineson17} we compiled pressure ratios for the FRII sub-sample of \citetalias{ineson15}, using X-ray inverse Compton measurements to determine the internal pressures, and environmental pressure profiles to determine the external pressures. Here we use this FRII sample, consisting of 37 objects, to compare with a sample of FRIs. As discussed above, we assume no non-radiating particles are present, as such a contribution is not required by the source dynamics (i.e. we assume $\kappa = U_{\rm p}$/$U_{\rm e} = 0$, where $U_{\rm p}$ and $U_{\rm e}$ are the energy densities of protons and electrons, respectively). To enable a comparison between FRI and FRII radio galaxies, we have compiled pressure ratios for the FRI environments subsample of \citetalias{ineson15}, consisting of 27 objects, with the results listed in Table~\ref{tab:fr1sample}. We used environmental pressure profiles to determine the external pressures, identically to the method used for the FRIIs. The internal pressure profiles could not be determined in the same way, as the magnetic field strength is not known (X-ray inverse Compton emission is not detected from the FRI sources: see Section~\ref{sec:ic}, below). We therefore considered two scenarios for the FRI internal pressures: (1) we assumed equipartition of energy density between radiating particles and magnetic field, assuming $\kappa = 0$ as has typically been done for previous FRI studies \citep[e.g.][]{croston03,croston08}; and (2) we assumed that the lobe magnetic field strengths of the FRIs have a similar distribution to the FRIIs, i.e. the magnetic field strength $B$ is typically $\sim 0.4 B_{\rm eq}$, where $B_{\rm eq}$ is the equipartition field strength assuming $\kappa=0$. We note that scenario (2) is already ruled out in some objects by X-ray inverse Compton limits (see Section~\ref{sec:ic}); however, we wanted in this study to consider the population as a whole. For both scenarios the internal pressures are calculated using the {\sc synch} code of \citet{hardcastle98}, with the same electron energy distribution assumptions as for the FRII results of \citetalias{ineson17}. The lobes are modelled as having uniform internal pressure. Lobe tip and mid-point distances are determined as in \citetalias{ineson17}, using the 3-$\sigma$ outer contour of radio emission.

\begin{table*}\centering
\caption{Pressure measurements for the FRI sub-sample. The second and third columns are the internal pressures for scenarios 1 and 2, respectively, as described in the text, columns 4 and 6 are the radial distances at which the comparisons are made, columns 5 and 7 are the external pressures at those distances, and column 8 indicates radio morphology (L = lobed, T = tailed, A = ambiguous morphology).}
\label{tab:fr1sample}
\begin{tabular}{lccccccc}\hline
Source&$P_{\rm int,Beq}$&$P_{\rm int,0.4B}$&Lobe-tip&$P_{\rm ext}$&Mid-lobe&$P_{\rm ext}$\\
&&&distance&at lobe-tip&distance&at mid-lobe&Morphology\\
&(${10}^{-13}$~Pa)&(${10}^{-13}$~Pa)&(kpc)&(${10}^{-13}$~Pa)&(kpc)&(${10}^{-13}$~Pa)\\
\hline
3C31&0.0367&0.107&411&$0.357^{+0.076}_{-0.023}$&205&$0.713^{+0.073}_{-0.011}$&T\\
3C66B&1.21&2.98&152&$3.46^{+0.04}_{-0.07}$&79&$5.99^{+0.11}_{-0.11}$&A\\
3C76.1&0.891&2.19&64&$0.704^{+0.326}_{-0.199}$&32&$1.76^{+0.52}_{-0.37}$&L\\
3C293&1.13&2.78&110&$0.101^{+0.040}_{-0.131}$&63&$0.349^{+0.146}_{-0.358}$&L\\
3C296&0.517&1.27&111&$1.85^{+0.14}_{-0.08}$&56&$2.72^{+0.15}_{-0.14}$&L\\
3C310&0.509&1.25&175&$5.22^{+0.10}_{-0.10}$&88&$13.5^{+0.3}_{-0.3}$&L\\
3C338&1.96&482&37&$177^{+0}_{-0}$&20&$268^{+0}_{-0}$&A\\
3C346&30.5&74.8&22&$68.1^{+5.6}_{-6.6}$&11&$123^{+10}_{-13}$&A\\
3C386&1.02&2.50&51&$1.36^{+0.16}_{-0.16}$&32&$3.00^{+0.29}_{-0.23}$&L\\
3C442A&0.231&0.567&183&$0.616^{+0.014}_{-0.012}$&91&$3.59^{+0.08}_{-0.08}$&A\\
3C449&0.102&0.250&269&$1.57^{+0.05}_{-0.08}$&134&$2.38^{+0.05}_{-0.08}$&T\\
3C465&0.669&1.64&236&$9.02^{+0.15}_{-0.15}$&127&$15.7^{+0.2}_{-0.2}$&T\\
NGC6109&0.132&0.325&521&$0.103^{+0.013}_{-0.091}$&261&$0.367^{+0.090}_{-0.224}$&T\\
NGC6251&0.0249&0.0610&998&$0.111^{+0.014}_{-0.006}$&583&$0.195^{+0.011}_{-0.006}$&T\\
NGC7385&4.63&11.4&30&$<$3.45&15&$<$1.28&A\\
0620-52&3.43&8.37&67&$46.8^{+1.7}_{-1.8}$&36&$24.7^{+1.3}_{-1.5}$&T\\
0625-35&1.29&3.16&92&$16.6^{+0.7}_{-0.7}$&49&$21.5^{+1.6}_{-0.9}$&T\\
0625-53&10.4&25.4&57&$76.7^{+1.9}_{-1.9}$&31&$89.9^{+3.3}_{-3.0}$&T\\
0915-11&15.9&38.9&78&$81.9^{+0.2}_{-0.2}$&41&$205^{+0}_{-0}$&T\\
1648+05&4.41&10.8&264&$25.1^{+0.2}_{-0.2}$&132&$70.4^{+0.6}_{-0.6}$&L\\
1839-48&5.48&13.5&88&$39.3^{+1.3}_{-1.2}$&44&$46.3^{+2.0}_{-1.7}$&A\\
1954-55&14.3&35.1&53&$22.4^{+0.5}_{-0.6}$&26&$33.8^{+1.7}_{-1.8}$&T\\
TOOT 1301+3658&0.908&2.23&93&$<$4.22&46&$<$9.64&L\\
TOOT 1255+3556&0.548&1.34&131&$1.61^{+0.54}_{-1.61}$&65&$5.52^{+3.39}_{-1.99}$&A\\
TOOT 1626+4523&0.422&1.03&217&$4.73^{+1.16}_{-0.77}$&108&$8.72^{+1.59}_{-1.19}$&A\\
TOOT 1630+4534&0.689&1.69&116&$<$5.28&58&$<$12.8&L\\
TOOT 1307+3639&1.52&3.74&78&$2.53^{+0.96}_{-2.53}$&38&$9.55^{+3.56}_{-9.41}$&L\\
\hline\end{tabular}\end{table*}

\begin{figure*}
  \includegraphics[width=15.0cm]{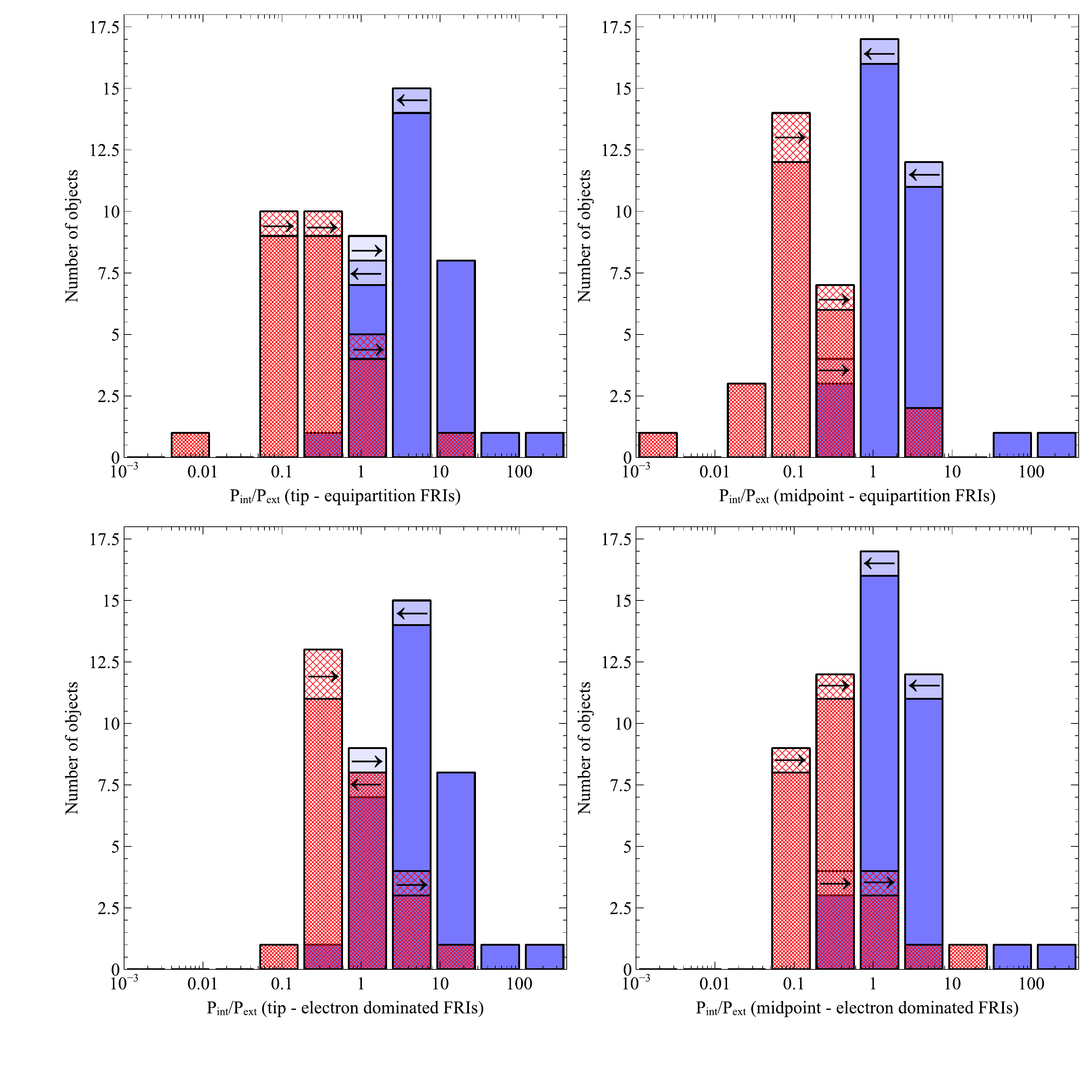}
\caption{Pressure ratio comparisons for the FRI (red hatched) and FRII (blue solid) sub-samples of \citetalias{ineson15}. The top row shows a histogram of the ratio of internal to external pressure for the FRI and FRII subsamples with the FRII ratios calculated from the X-ray inverse-Compton measurements, and assuming $\kappa=0$, as described in \citetalias{ineson15}, and the FRI ratios calculated assuming equipartition and $\kappa=0$. The bottom row shows the same FRII ratios, but for the FRI estimates makes the assumption that they have the typical departure from equipartition measured for the FRIIs ($B = 0.4 B_{\rm eq}$), as discussed in the text. A small number of FRIs and FRIIs with limits rather than measurements of the pressure ratio are included and indicated by the arrow symbols.}
\label{fig:ratios}
\end{figure*}

Fig.~\ref{fig:ratios} shows histograms comparing the ratio of $P_{\rm int}$/$P_{\rm ext}$ for the FRI and FRII sub-samples of I15, including limits ($\sim 10$ percent of each sample). The top panels assume scenario (1) for the FRIs, and the bottom panels assume scenario (2), with the left panels showing the lobe-tip pressure ratios, and the right panels the midpoint ratios. For the equipartition FRI scenario (1), the FRIs are typically substantially {\it under-}pressured at both the lobe tips and the midpoints. For the FRII-like magnetic field strength scenario (2), the FRIs remain substantially under-pressured at the lobe tip; they are closer to pressure balance in this scenario, but 18/24 measured midpoint pressure ratios $P_{\rm int}$/$P_{\rm ext}$ are less than 1, and 16/24 measured lobe-tip ratios are than 1. Our results therefore demonstrate conclusively that either a departure from equipartition significantly larger than that of the FRIIs, or a substantial proton contribution is required for the FRI radio-lobe population, in agreement with previous work \citep[e.g.][]{croston08,croston14}. The small number of pressure ratio limits, which occur mainly in situations where only a limit on the external pressure is available, are indicated by the arrows symbols in Fig.~\ref{fig:ratios}. There are also two FRIIs in the sample for which both internal and external pressures are limits, so that the pressure ratio is unconstrained -- we exclude those in our comparison. The relatively small number of limits does not affect the main conclusion of a systematic difference in the FRI and FRII distributions for both the midpoint and tip distributions. We cannot use survival analysis to compare the subsample distributions, as there are both upper and lower limits included; however, if the small number of limits are excluded then for all four comparisons shown in Fig.~\ref{fig:ratios}, a Wilcoxon-Mann-Whitney test finds that the null hypothesis of the observed pressure ratio distributions for the two subsamples resulting from the same underlying distribution has a probability $<0.1$ per cent. In Table~\ref{tab:meds}, we list the limits on the median pressure ratios for each subsample and scenario, taking into account the small number of non-detections, which confirm the differences between the FRI and FRII energetics in all four cases.

In addition to considering the FRI objects as a single class as shown in Fig.~\ref{fig:ratios}, we also split the sample into tailed and lobed morphology subclasses \citep[e.g.][]{croston08} -- these classifications are listed in the final column of Table~\ref{tab:fr1sample}. We then compared the pressure ratios of each FRI subclass separately with those of the FRII population using a Wilcoxon-Mann-Whitney test. For each subclasses the null hypothesis that it has the same underlying distribution of pressure as the FRIIs could also be rejected ($<0.1$ per cent probability in each case). We discuss the relationship between particle content and FRI morphology further in Section~\ref{sec:morph}.

\subsection{Proton content in FRI and FRII radio lobes}
The comparison of FRI and FRII pressure ratios in Fig.~\ref{fig:ratios} strongly suggests a difference in the internal energetics for the two populations. We have previously argued that entrainment of protons from the surrounding gas is the favoured explanation for the ``missing'' pressure in FRI radio galaxies \citep[e.g.][]{croston08,croston14}. We now consider whether our results can rule out FRI and FRII radio galaxies typically possessing similar levels of proton content.

In the case of FRI equipartition magnetic fields (scenario 1), the results shown in Fig.~\ref{fig:ratios} indicate that a median energy ratio of non-radiating to radiating particles of $U_{\rm p}$/$U_{\rm e} \sim 9$ would be required to enable the FRIs in this sample to be in pressure balance at their midpoints. If the FRIIs were to have a similar level of proton content this would lead to median pressure ratios at the lobe midpoint of $P_{\rm int}$/$P_{\rm ext} \sim 16$, with seven systems being over-pressured at their midpoints by a factor $> 40$. We have recently found \citep{croston17} that FRII internal pressures calculated assuming no protons and $B = 0.4B_{\rm eq}$ are a reliable predictor of the external pressure at the lobe midpoint, consistent with the results of hydrodynamical simulations finding approximate pressure balance at FRII midpoints \citep[e.g.][]{hardcastle13}. This result would have to be due to a strong coincidence if in fact the internal pressures are more than an order of magnitude higher, and dominated by a proton contribution. We interpret this as a strong argument against a model in which all radio galaxies, both FRI and FRII, are energetically dominated by non-radiating particles.

\begin{figure}
  \includegraphics[width=7.0cm]{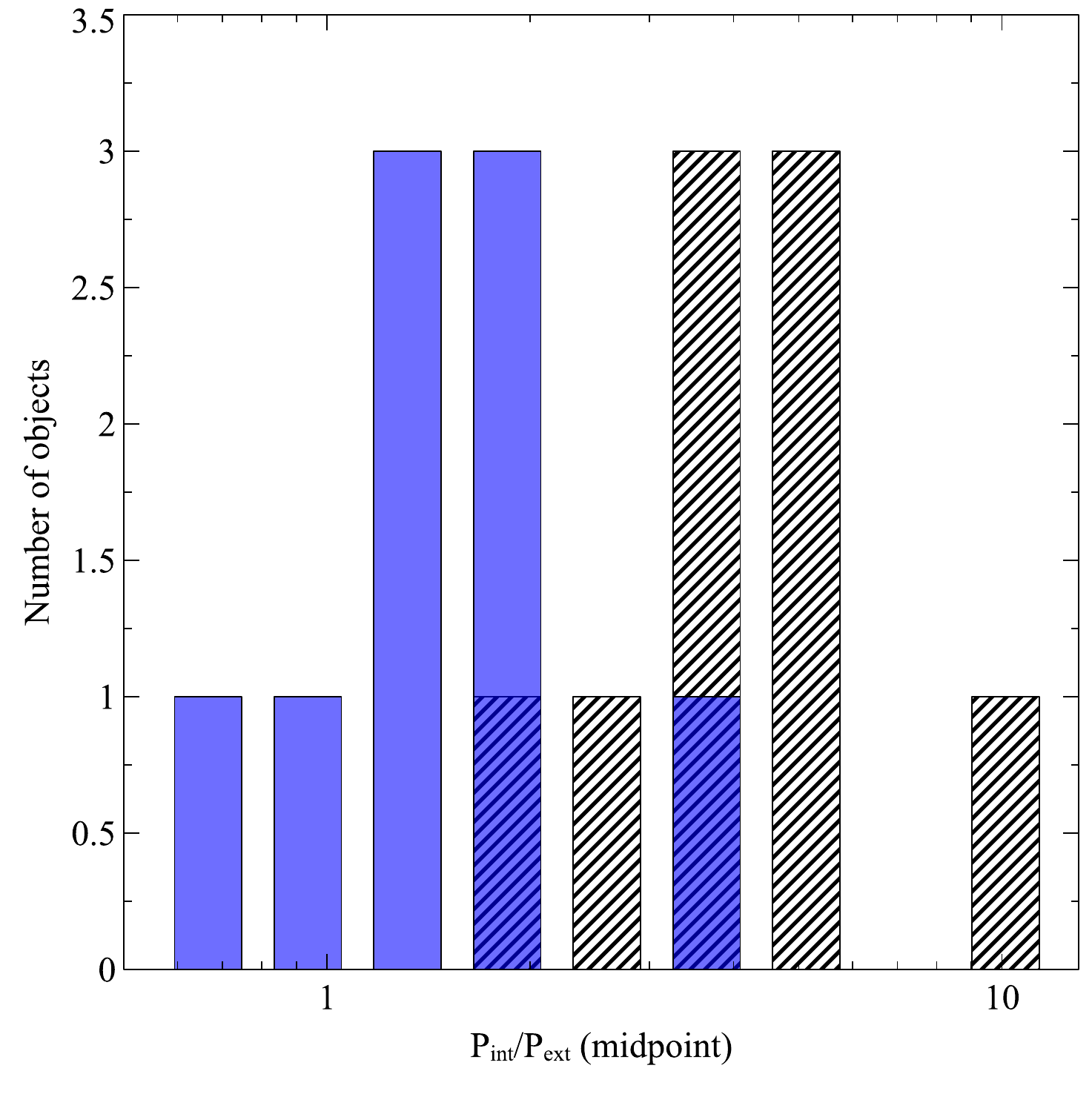}
\caption{A comparison of FRII pressure ratios at the ``pinch
    points'' where the lobe material appears compressed by the
    environment, for the observed IC internal pressures (blue)
    and for the assumption of $\kappa = 3.4$ (black diagonal lines), the typical value
    required to bring the FRIs into pressure balance at their midpoints in scenario 2 (FRII-like $B$ ratios).}
\label{fig:pinch}
\end{figure}

It might be expected, however, that if the FRIs and FRIIs have similar proton content, they would also have a similar energy balance between particles and magnetic field, as in  scenario 2, in which the FRIs also have sub-equipartition magnetic fields ($B \sim 0.4 B_{\rm eq}$), similar to the FRIIs. For scenario 2, bringing the FRIs into pressure balance at their midpoints would require a typical energy ratio of non-radiating to radiating particles of 3.4, i.e. a relatively modest increase in pressure ratio. Could such a non-radiating particle population be accommodated in the FRII population as well? For some of the FRIIs we are able to identify particular locations where the lobes must be close to pressure balance, which provide the most stringent constraints on what internal pressures are physically plausible. We identified a sub-sample of seven FRIIs where the lobe/bridge material extending back towards the nucleus narrows or is pinched off (these are 3C\,33, 3C\,35, 3C\,219, 3C\,452, PKS\,0043$-$42, PKS\,0349$-$27, and PKS\,1559$+$02). The morphology of these sources indicates that the pressure of the surrounding gas is pinching the radio cocoon inwards, and hence they should not be significantly over-pressured at those points \citep[e.g.][]{hardcastle13,hardcastle14}. Fig.~\ref{fig:pinch} shows the pressure ratios at the pinch points for these seven FRIIs (plotting the two lobes separately for two objects where the pinch point occurs at a different distance on each side) using the inverse-Compton measured pressure ratios (assuming no protons, shown in blue), and with the required non-radiating contribution of 3.4 times higher (in black hatching) that would allow the FRI and FRII particle content to be the same (on average). We find that the $\kappa=0$ (``observed'') pressures lead to ratios
distributed close to unity, as suggested by the source morphology, whereas the scaled up ratios would mean all of the lobes are still over-pressured at the locations where pinching appears to be taking place. Hence lobe morphologies appear to rule out the scaled up pressures for these seven objects ($\sim 25$ per cent of the FRII sample).

\subsection{X-ray inverse-Compton limits for the FRI subsample}
\label{sec:ic}
A further constraint on the magnetic field strengths for the FRI radio galaxies comes from X-ray inverse-Compton upper limits. If the FRIs had similar $B$/$B_{\rm eq}$ ratios to the FRIIs, then in some cases this would lead to a prediction of detectable X-ray IC emission (with the predictions being even higher if electron domination were to provide all of the pressure required by dynamical considerations as is the case for the FRIIs). We have previously shown that X-ray inverse-Compton limits rule out such models in several individual cases \citep{hardcastle98,croston03,hardcastle10,croston14}. For a subset of our FRI sample where the predicted levels are in principle detectable, we obtained upper limits on the lobe X-ray inverse-Compton emission, using the same method and assumptions as for the FRII inverse-Compton measurements of \citetalias{ineson17}. We compared these with the predictions for (1) the scenario where $B=0.4B_{\rm eq}$, shown in the lower panels of Fig.~\ref{fig:ratios}, and (2) the scenario where electrons provide all of the pressure required for mid-lobe pressure balance. Table~\ref{tab:ic} reports the results of this comparison: we find that both scenarios can be ruled out for four FRIs in our sample, while scenario (2) can also be ruled out for a further three FRIs. It is also worth noting that in models where radio lobes contain a significant proton population the relative similarity of IC-measured magnetic field strengths and the equipartition value for $\kappa=0$ must be a coincidence: approximate equipartition between the energy density of all particles and of the magnetic field in such a scenario would require $B>B_{\rm eq}$, i.e. a magnetic field strength {\it higher} than the no-proton equipartition value, rather than lower as observed for the FRIIs. The entrainment model of \citet{croston14} is an example of a situation where $B>B_{\rm eq}$ for the assumption of $\kappa=0$.

\begin{table}
\caption{Limits on the median pressure ratios for the subsamples shown in Fig.~\ref{fig:ratios}. The FRI medians are the {\it upper limit} to the median value assuming that the three lower limits could take values at the extreme high end, and the FRII medians are the {\it lower limits}, assuming that the two upper limit pressure ratios could take values at the extreme low end of the FRII distribution. Hence the FRI/II comparisons are based on the most conservative assumption of how similar the median pressure ratios could be. We also include the median ratios separately for lobed and tailed FRIs for the $B_{\rm eq}$ assumption, confirming a possible difference at the tip, but no evidence of a difference at the midpoint -- for this comparison the medians are not limits, as there are no limits in the tailed sample, and both limits in the lobed sample are lower limits so that the median rather than a limit on the median can be determined.}
\label{tab:meds}
\begin{tabular}{lccc}\hline
Sample&Position&$B$ assumption&Median $P_{\rm int}$/$P_{\rm ext}$\\
\hline
FRI&mid&$B_{\rm eq}$&$<0.13$\\
FRI&mid&$0.4 B_{\rm eq}$&$<0.34$\\
FRII&mid&-&$>1.4$\\
FRI&tip&$B_{\rm eq}$&$<0.28$\\
FRI&tip&$0.4 B_{\rm eq}$&$<0.84$\\
FRII&tip&-&$>3.9$\\
tailed FRI&tip&$B_{\rm eq}$&0.14\\
tailed FRI&mid&$B_{\rm eq}$&0.12\\
lobed FRI&tip&$B_{\rm eq}$&0.60\\
lobed FRI&mid&$B_{\rm eq}$&0.19\\
\hline\end{tabular}
\end{table}

We therefore conclude that FRI and FRII radio galaxies appear to have systematically different particle content and energy division between particles and magnetic field. Our comparison of the FRI and FRII populations shows that the average non-radiating particle contribution required by FRIs could be relatively modest if they have sub-equipartition magnetic fields similar to the FRIIs; however, IC limits rule out such magnetic fields in a subset of the sample, while FRII morphologies and the requirement of a conspiracy to enable the no-proton pressure estimates to coincide with the environmental mid-lobe pressures argue against the FRII population containing a similar level of non-radiating particles as required for the FRIs. Hence it appears that the magnetic field ratios ($B$/$B_{\rm eq}$) in the FRI radio galaxies are different from those of the FRIIs, and the true proton content may be much higher in many FRIs, as we have argued elsewhere \citep[e.g.][]{croston14,heesen17}.

\subsection{Particle content and FRI morphology}
\label{sec:morph}

\begin{figure}
\includegraphics[width=7.0cm]{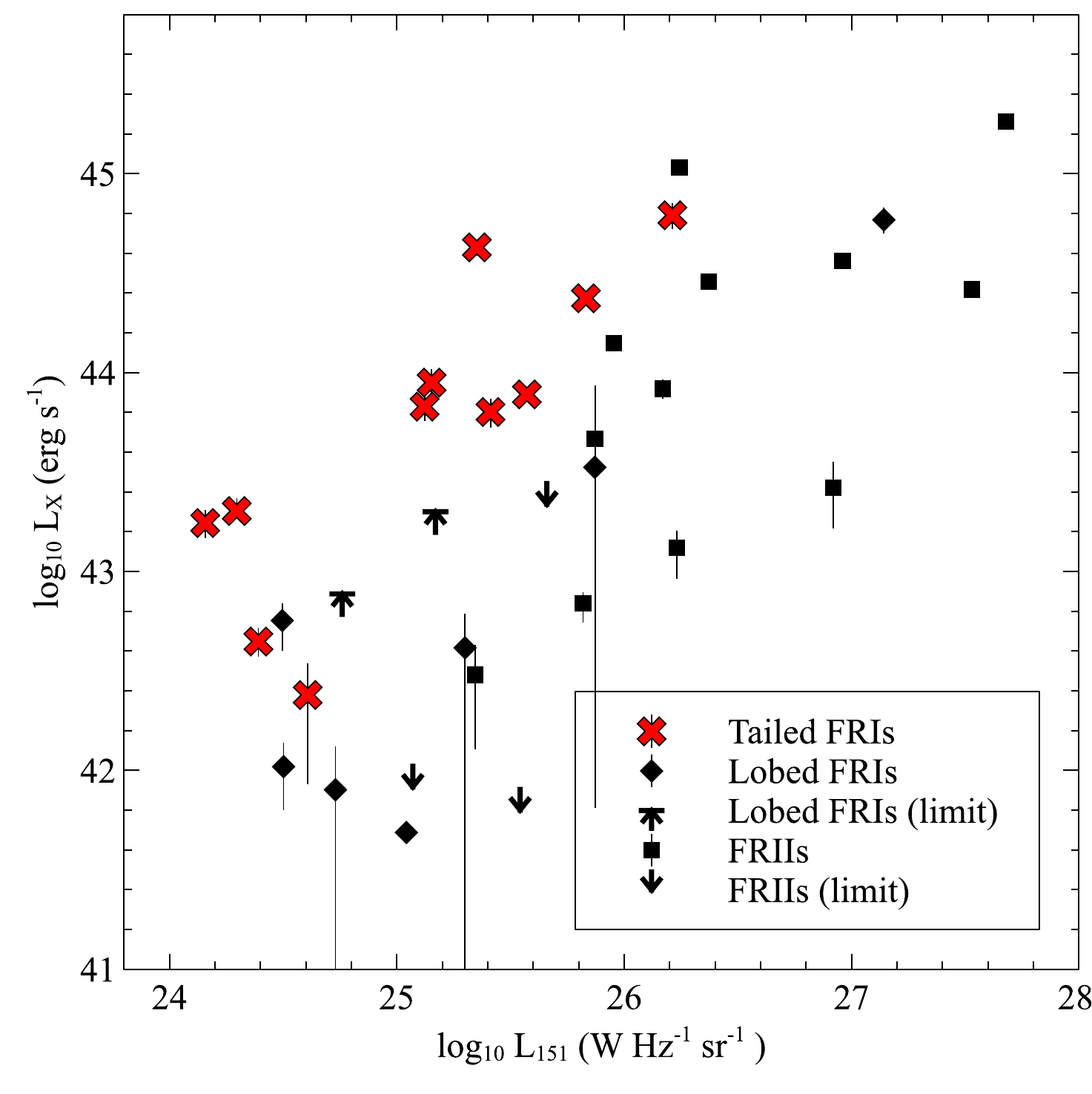}
\caption{The X-ray environment -- radio luminosity relation for
    LERGs, from \citet{ineson15}, broken down by radio morphology. A small number of upper limits on the X-ray environmental luminosity are present, as indicated in the legend.}
\label{fig:morphenv}
\end{figure}

While in the previous section we considered a constant ratio of non-radiating to radiating particle energy for all FRIs, observations demonstrate clearly that this factor must vary significantly throughout the FRI population \citep[e.g.][]{birzan08,dunn04,croston08}. In the study of \citet{croston08}, we found a relationship between $P_{\rm eq}$/$P_{\rm ext}$ and FRI morphology, in the sense that tailed FRIs appear to require higher values of $\kappa$ than lobed FRIs.
Some further evidence for the importance of radio morphology as a
diagnostic of radio-lobe internal conditions comes from our investigations of the relationship between radio properties and environment (\citetalias{ineson15}). For the low-excitation radio galaxies in our sample, we found a strong relationship between $L_{R}$ and environmental richness, characterized by the cluster X-ray luminosity, $L_{X}$, albeit with large scatter. In seeking to understand the source of scatter in this relation we have discovered that radio morphology appears to play a role in locating FRIs in the $L_{R}$--$L_{X}$ plane. Fig.~\ref{fig:morphenv} shows the LERG $L_{R}$--$L_{X}$ relation, with the FRIs broken down by morphology. Tailed FRIs typically lie in richer environments than the general population at the same radio luminosity. If the underlying relationship driving the observed $L_{R}$--$L_{X}$ correlation is a relationship between jet power and environment (\citetalias{ineson15}), the morphological effect shown in Fig.~\ref{fig:morphenv} could be explained by the role of particle content: if the tailed sources are effectively ``under-luminous'' for their jet power, due to much of the jet power being carried by non-radiating particles, then at the same radio luminosity they are more powerful than their lobed counterparts. \citet{croston08} argued that the jets of tailed sources are in direct contact with the external medium on 10 - 100 kpc scales where entrainment would be required \citep[e.g.][]{croston14}, while the jets of lobed FRIs are embedded within cocoons and thus cannot entrain from the external medium (though they can entrain stellar material on small scales: e.g. \citealt{laing02,hardcastle07,wykes15}), so that higher levels of proton content may be expected in tailed FRIs. Another contributing factor may be mixing, which becomes more important as buoyancy effects raise lobe plasma outwards -- this could be particularly relevant for the most extreme radio bubbles and ``ghost''
cavities in clusters.

\begin{figure*}
\includegraphics[width=15.0cm]{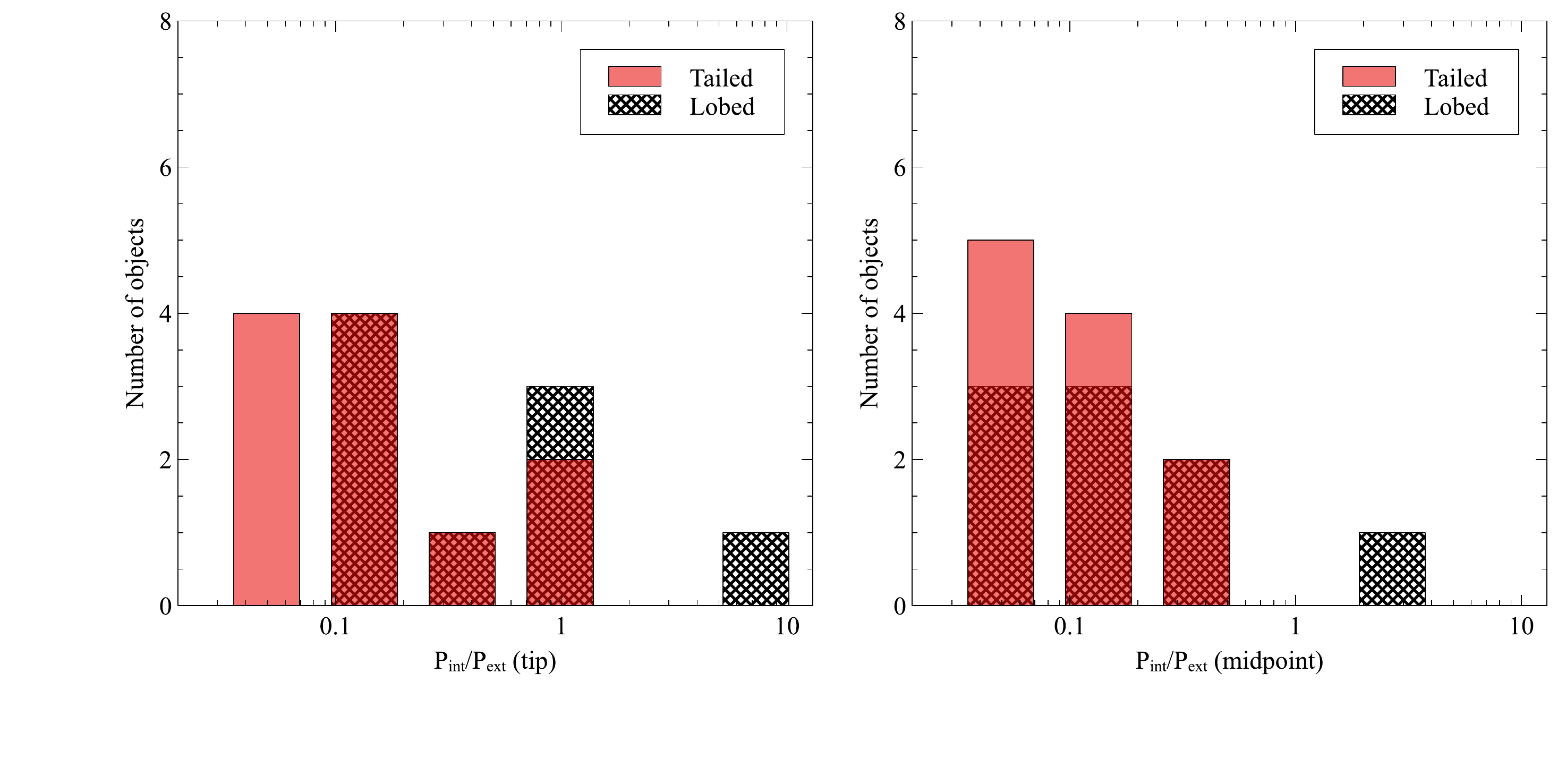}
\caption{A comparison of FRI pressure ratios for tailed and lobed FRIs, for the assumption of $B=B_{\rm eq}$ and $\kappa=0$. Pressure ratio limits are omitted.}
\label{fig:morphcomp}
\end{figure*}

We investigated whether tailed sources in this, more representative, sample of objects may have systematically higher proton content than lobed FRIs. Fig.~\ref{fig:morphcomp} compares the FRI pressure ratios for the lobed and tailed sub-samples of our FRI sample. While there is a hint of a difference between the two samples, a Wilcoxon-Mann-Whitney test cannot rule out that the tailed and lobed sources are drawn from the same parent population, and as we note in Section~\ref{sec:comp}, both subsamples have a systematically different distribution of pressure ratios to the FRIIs. We also determined the median pressure ratios for the lobed and tailed sub-samples, which are listed in Table~\ref{tab:meds}. The medians are quite different at the lobe tip, but indistinguishable at the lobe midpoint, consistent with the plots in Fig.~\ref{fig:morphcomp}. The difference in lobe-tip medians could simply reflect the fact that we are using a single internal pressure estimate, which is a less valid assumption for tailed sources where the internal pressure is expected to decrease with distance, and so we cannot conclude from this that there is a particle-content difference between the two sub-classes of FRI. Any difference in particle content between the two sub-populations of FRI therefore appears to be subtle and cannot be the primary explanation for the location of the tailed FRIs in Fig.~\ref{fig:morphenv}. An alternative explanation is that the radio morphology for jets of similar density/pressure will evolve differently in a richer environment compared to a poorer one. The combined effects of a denser inner environment and stronger effects of buoyancy may preferentially cause tail-like morphologies in richer environments \citep[e.g.][]{hardcastle99}. Our results do not allow us to distinguish between these two explanations. 

We have recently shown that for FRII radio galaxies it is possible to predict the environment X-ray luminosity reliably from the radio-estimated internal pressure \citep{croston17}. The method described in that work is not applicable to FRI radio galaxies due to their different energetics. Fig.~\ref{fig:morphenv} suggests that, instead, the combination of radio luminosity and morphology may enable reliable environmental predictions for tailed FRI sources (size could be an additional relevant factor).

\section{Implications for jet power relations}
\label{sec:qjet}
If, as concluded above, the particle content of
radio galaxies can vary substantially (from an energetically
negligible proton content to proton-to-electron energy ratios of
$>100$), this has implications for estimates of jet power from radio
observable quantities that trace only the leptons and magnetic
field. Indeed \citet{willott99} include particle content ($\kappa$) as an unknown factor in their widely used relation between jet power ($Q_{\rm jet}$) and radio luminosity ($L_{\rm R}$). The value of $\kappa$ has also been assumed to be one of the factors introducing substantial scatter into the observed relations between radio luminosity and jet power estimated from X-ray cavities \citep[e.g.][]{birzan08}. The key implication of the results that we present in the previous Sections is that varying particle content is likely to introduce a systematic bias for certain sub-populations, rather than simply random deviations from some `canonical' $Q_{\rm  jet}$--$L_{\rm R}$ relation. As the demographics of the radio-loud AGN population (e.g. the breakdown of the luminosity function into FRI/II and LERG/HERG classes) are known to evolve with
redshift \citep[e.g.][]{williams16,best14}, such systematic effects mean that a $Q_{\rm jet}$--$L_{\rm R}$ relation derived for local FRIs may lead to systematically incorrect results if extrapolated to high redshift jet populations.

We therefore investigated the implications of the systematic difference in FRI and FRII particle content on the scaling relations between $Q_{\rm jet}$ and $L_{R}$. In \citetalias{ineson17} we presented new jet power estimates for the FRII radio galaxy population based on our X-ray inverse Compton measurements. This jet power method involves a direct observational estimate of lobe expansion speed from lobe over-pressuring, which has not previously been available for samples of radio galaxies, and results in jet power estimates consistent with other methods \citep[e.g.][]{daly12,godfrey13}. We demonstrated (\citetalias{ineson17}) that the relation we observe for our FRII sample is not driven by distance dependence -- a concern raised for some other jet power studies by \citet{godfrey16}.
 
 \begin{figure*}
  \includegraphics[width=7.0cm]{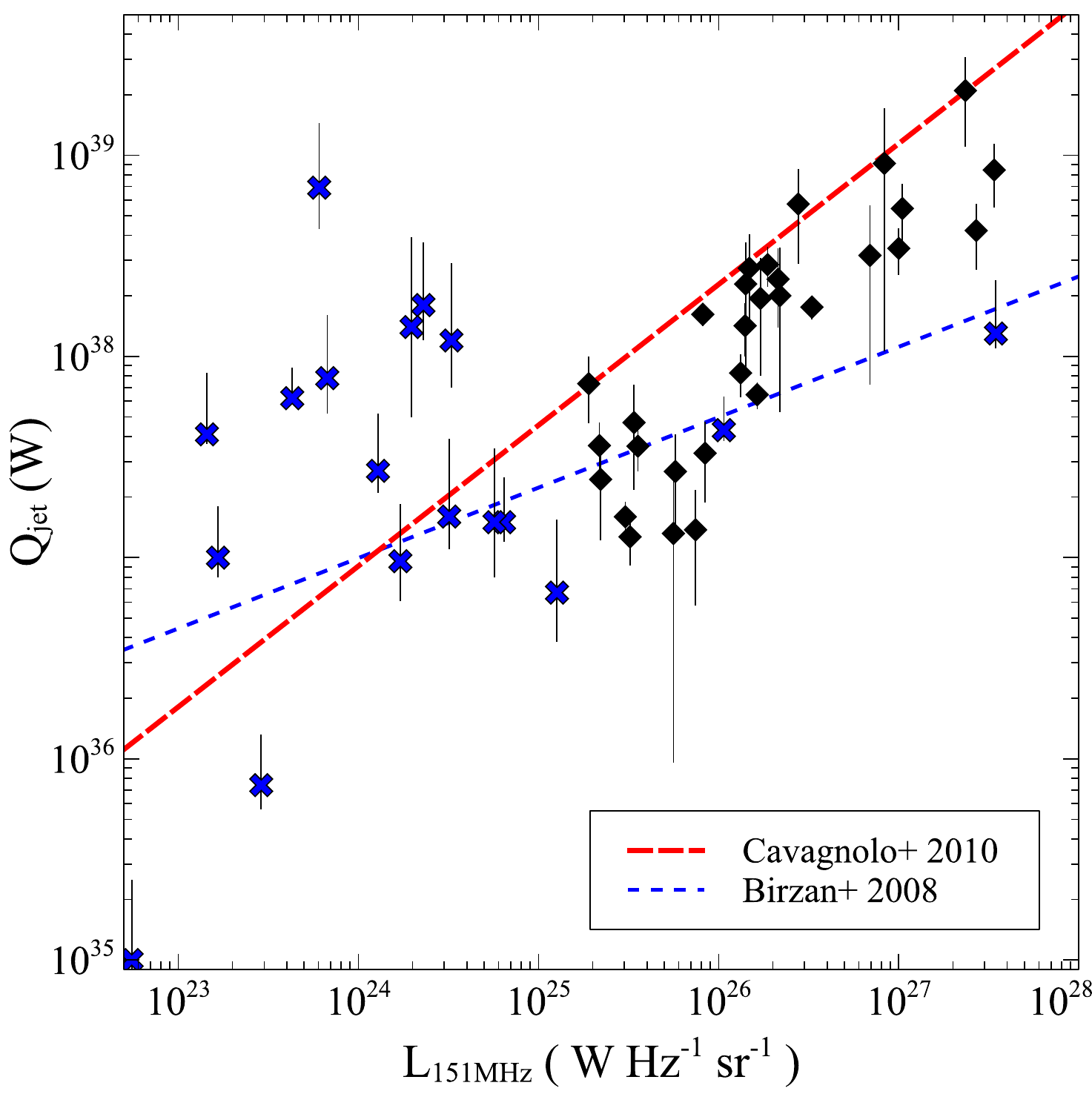}
\includegraphics[width=7.0cm]{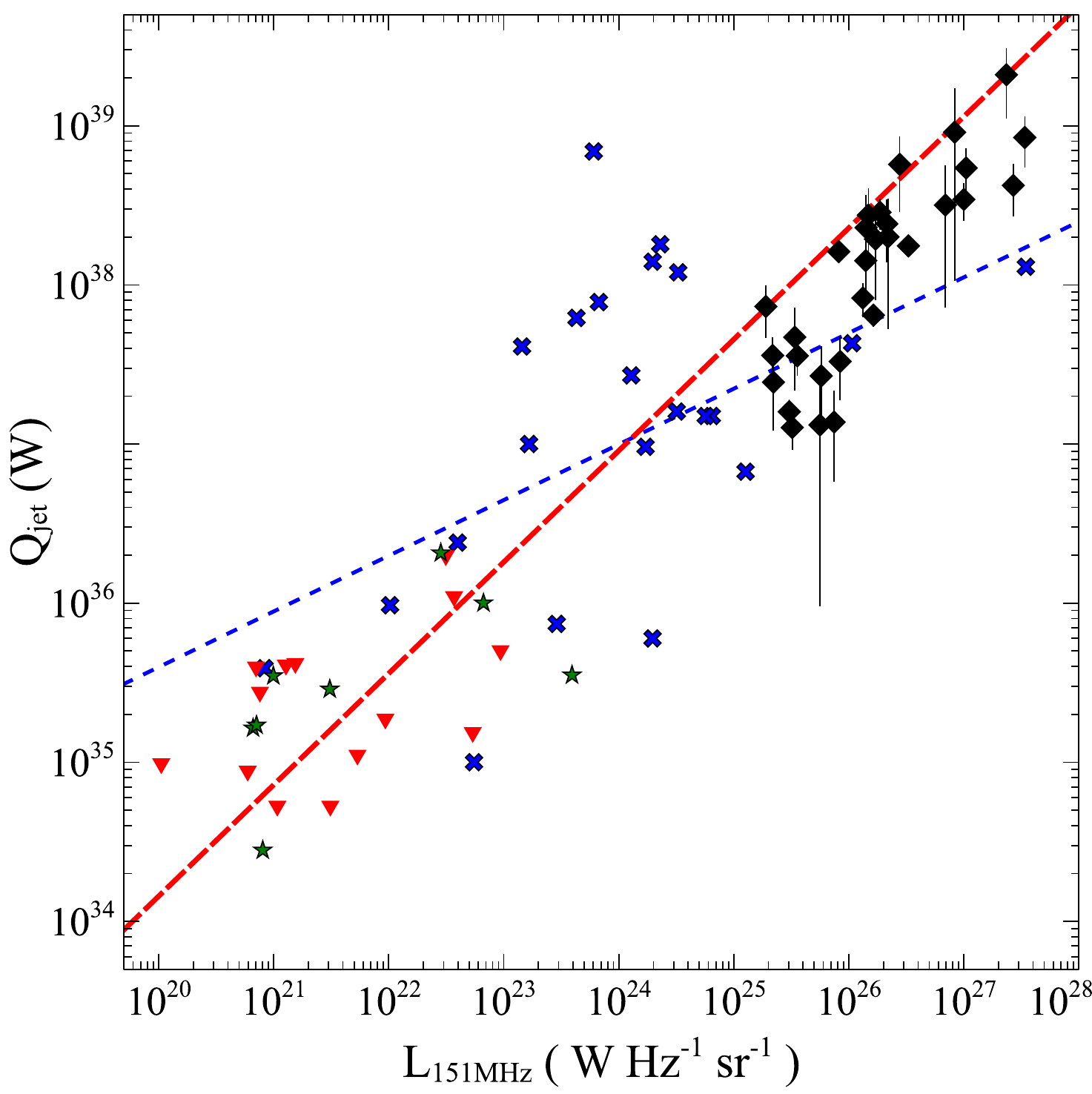}
\caption{The relationship between 151-MHz radio luminosity and jet
    power; Left: The \citetalias{ineson17} FRII sample (black diamonds), compared with the sample of \citet{birzan08} (blue crosses), with the average $Q_{\rm jet}$--$L_{\rm R}$ relation of \citet{cavagnolo10} (red) and the relation of \citet{birzan08} (blue) overplotted; Right: the same comparison but with jet powers from the FRI cavity samples of \citet{cavagnolo10} and \citet{osullivan11}(green stars). We omit error bars for the cavity samples in the right-hand panel for clarity.}
\label{fig:qjet}
\end{figure*}
 
We do not have dynamical $Q_{\rm jet}$ estimates for our FRI sub-sample, as they are not X-ray-IC-detected. We therefore compared our FRII jet powers with previously published work using X-ray cavity energetics to enable us to make an assessment of the implications of our particle content results for the inference of jet power from large radio-loud AGN
  samples. Fig.~\ref{fig:qjet} shows a comparison of our results with the
previously published jet power measurements and $Q_{\rm jet}$--$L_{R}$ scaling relations of \citet{cavagnolo10}, \citet{birzan08} and \citet{osullivan11} (with the cavity samples consisting almost exclusively of FRI morphology sources -- an exception is Cygnus A, which is the blue cross to the far right of the plots). It is immediately noticeable from the left-hand panel that the range of inferred FRII jet powers is broadly similar to that of the \citet{birzan08} cluster cavity sources, whose radio luminosities are considerably lower. The average jet-power relation from \citet{cavagnolo10} over-predicts the FRII jet powers, while the \citet{birzan08} relation, which is very flat (and poorly constrained due to the small range in radio luminosity and large scatter in jet power) on average under-predicts them. It is important to note that the choice of jet-power relation can lead to an order of magnitude uncertainty in jet power at FRII radio luminosities. It is also worth emphasizing that the existence of both FRI and FRII radio morphologies over a similar range in jet power, as shown in the left-hand panel of Fig.~\ref{fig:qjet} is not unexpected: it is consistent with jets of a similar power being preferentially disrupted to form an FRI morphology in richer environments, while remaining undisrupted and FRII-like in poorer environments. The median cluster X-ray luminosity for the sample of \citet{birzan08} is $\sim 3 \times 10^{44}$ erg s$^{-1}$ \citep{rafferty06}, while the median FRII environment from \citetalias{ineson17} is $\sim 10^{43}$ erg s$^{-1}$. 

Taking the \citet{cavagnolo10} relations, which are based on observations spanning a wider range in radio luminosity, the observed offset is what would be expected if the radio luminosity of (large, currently active) FRIIs is a more direct tracer of jet power than for the FRIs, while the FRIs contain a significant proton contribution. The right-hand panel of Fig.~\ref{fig:qjet} suggests, however, an alternative interpretation in which a subset of the (typically rich cluster hosted) sources in the \citet{birzan08} sample are particularly extreme. \citet{godfrey13} point out that an even larger offset between FRII and FRIs is predicted, based on the range of $\kappa$ values inferred for FRIs (which are particularly extreme for the cluster cavity systems of \citealt{birzan08}); however, as discussed by \citetalias{ineson17}, the departure from minimum energy we find from the FRII inverse Compton observations would compensate partially for the difference in proton content. 

It is apparent from Fig.~\ref{fig:qjet} that barring a small number of extreme cluster cavity systems, a single relation could be fitted across eight orders of magnitude in radio luminosity, with relatively small scatter. However, we have chosen {\it not} to fit a relation to the full plotted sample, as we believe it would be misleading. Our plot is based on two very different types of sample selection, which are likely to be substantially different to the selection function of new surveys to which such a relation might be applied (e.g. LOFAR surveys will select for a broader range in source age and evolutionary stage than either of the two methods presented here). Since radio luminosity at fixed jet power is known to evolve throughout a source's lifetime, due to both the dynamics of the source itself and the effect of radiative losses (e.g. \citealt{hardcastle13} find up to two orders of magnitude variation in luminosity with source age for a fixed jet power, not accounting for radiative losses), it is clear that differences in radio selection function will lead to further systematic differences in the relationship between $Q_{\rm jet}$ and $L_{R}$. We would expect that any evolution of the relationship between cluster environment and jet power would also alter the $Q_{\rm jet}$--$L_{R}$ relation. The importance of both radiative losses and environment in driving very large scatter in this relationship is also highlighted in the recent modelling work by \citet{hardcastle18}.

\subsection{Systematics in the comparison of cavity and IC jet powers}
Given the use of two different jet power estimate methods in Fig.~\ref{fig:qjet}, it is important to consider any sources of systematic offset between the FRII and FRI results. The FRII jet power estimates assume no protons; increasing the FRII internal pressures by assuming that protons contribute a similar fraction of the total energy as for the FRIs would remove the FRI/II difference in Fig.~\ref{fig:qjet}. We have argued against this scenario in Section~\ref{sec:pcont}, and the purpose of the previous section is to explore the consequences of the inferred FRI/II particle content difference discussed in Section~\ref{sec:pcont} for the $Q_{\rm jet}$ relations. We also note again that the presence of FRI and FRII sources of similar jet powers is not unexpected given the systematically different environments of the samples compared in Fig.~\ref{fig:qjet}. In this Section we discuss other sources of systematic differences. 

The timescales assumed for the FRII calculation are based on the Mach numbers inferred from the pressure ratio comparison (see \citetalias{ineson17} for details), whereas timescales for the cavity samples assume expansion on a buoyant timescale. The cavity timescales are more likely to overestimate rather than underestimate the true timescale, as expansion in some stages of evolution will have been supersonic; hence, any systematic adjustment to account for this would {\it increase} the cavity $Q_{\rm jet}$ powers. For the FRIIs we have a direct estimate of the instantaneous expansion speed from the pressure comparisons, which provides a more direct timescale estimate. As discussed in \citetalias{ineson17}, we conclude that the instantaneous timescales inferred from the lobe-tip pressure ratio could underestimate the time-averaged expansion speed over the source lifetime (the quantity required for calculating jet power) by a small factor. If the cavity timescale estimates are correct, then to bring the FRIIs into agreement with the \citet{cavagnolo10} relation by assuming the timescales are overestimated would require increasing the typical lobe expansion speed by a factor 3 - 5, which is larger than the estimated correction for variation in expansion speed over the source lifetime from the simulations of \citet{hardcastle13}; however, this factor is not well constrained. While both timescale methods are uncertain, it is likely that any systematic offsets in the timescales will act in the same direction for both subsamples, and may be more significant for the cavity sources where no supersonic expansion is accounted for, which would increase the FRI/II offset in Fig.~\ref{fig:qjet}.

One of the main uncertainties in the energy estimates is the assumption about the ratio between lobe internal energy and energy transferred to the external medium. The results we plot for the cavity samples assume that the jet's total integrated energy output corresponds to $4PV$ (where $PV$ is determined from the observed cavity). This is likely to be an underestimate (as discussed by \citealt{birzan08} and \citealt{cavagnolo10}), as some additional energy will be transferred via shocks during some stages of the source's evolution. For the FRII sample we assume that the total energy transferred is twice the internal energy of the radio lobe,
as typically found in hydrodynamical simulations of FRII lobe evolution \citep[e.g.][]{hardcastle13,hardcastle14}, in other words that the total integrated energy output is $6PV$. Hence adjusting the cavity results to the same total energy relative to the lobe size would increase the inferred FRI jet powers, and so act to increase the FRI/II offset. To bring the FRIIs into line with the \citet{cavagnolo10} relation would require the lobes to transfer 6 -- 10 times their internal energy to the external medium -- there is no obvious physical motivation for such a large multiplier. 

We conclude that the most obvious sources of systematic uncertainty due to the two different jet power estimation methods, the possible overestimation of timescales and underestimation of total energy input for cavity sources, would both act mainly to {\it increase} the discrepancy between the FRII and FRI results, rather than reduce it. While there are substantial uncertainties for both methods of jet power estimation, our results demonstrate that a systematic difference in FRI and FRII particle content does introduce an offset in the relationship between radio luminosity and jet power, with FRIIs having a systematically lower ratio of jet power to radio luminosity. It is important to note that other factors such as source size, environment and age may be similarly important. Our work adds to the growing evidence that jet-power scaling relations need to be used with caution \citep[e.g][]{kokotanekov17,hardcastle18} -- more work is needed to understand how the $Q_{\rm jet}$--$L_{\rm R}$ relation evolves over source lifetimes for populations in a variety of environments.

\section{Particle content and accretion mode}
\label{sec:accmode}
Throughout this paper we have sub-divided our sample into the
traditional FRI and FRII classes, making the usual assumption that
this division is the result of a dynamical difference in jet evolution
\citep[e.g.][]{bicknell94,tchekhovskoy16}. However, it has been shown that there are two accretion modes operating within the radio-loud AGN population \citep[e.g.][]{hec07,heckman14}, and our earlier work (\citealt{ineson13} and \citetalias{ineson15}) demonstrated that the relationship between radio properties and cluster environment differs for LERGs and HERGs. It is therefore important to consider whether the difference in particle content we present here is firmly linked to FR class, or whether it could be related to accretion mode instead. 

\begin{table}\centering
\caption{X-ray inverse-Compton limits for FRIs with predicted detectable emission. Columns are: source, measured 1-keV lobe flux density, predicted 1-keV lobe flux density for $B=0.4B_{\rm eq}$, and predicted flux density for mid-lobe pressure balance and electron domination with no protons.}
\label{tab:ic}
\begin{tabular}{lccc}\hline
Source&$S_{\rm obs}$&$S_{\rm 0.4Beq}$&$S_{\rm P balance}$\\
&(nJy)&(nJy)&(nJy)\\
\hline
3C\,296&$<6.8$&5.1&10.6\\
PKS 0915$-$11&$<5.7$&4.9&25.9\\
3C\,388&$<10.9$&2.9&5.0\\
3C\,465&$<2.2$&10.7&108\\
3C\,310&$<1.4$&12.9&143\\
PKS 1648$+$05&$<38$&23.6&156\\
3C\,449&$<3.0$&16.0&160\\
NGC\,6251&$<3.1$&55.5&182\\
\hline\end{tabular}\end{table}

\begin{figure*}
  \includegraphics[width=14.0cm]{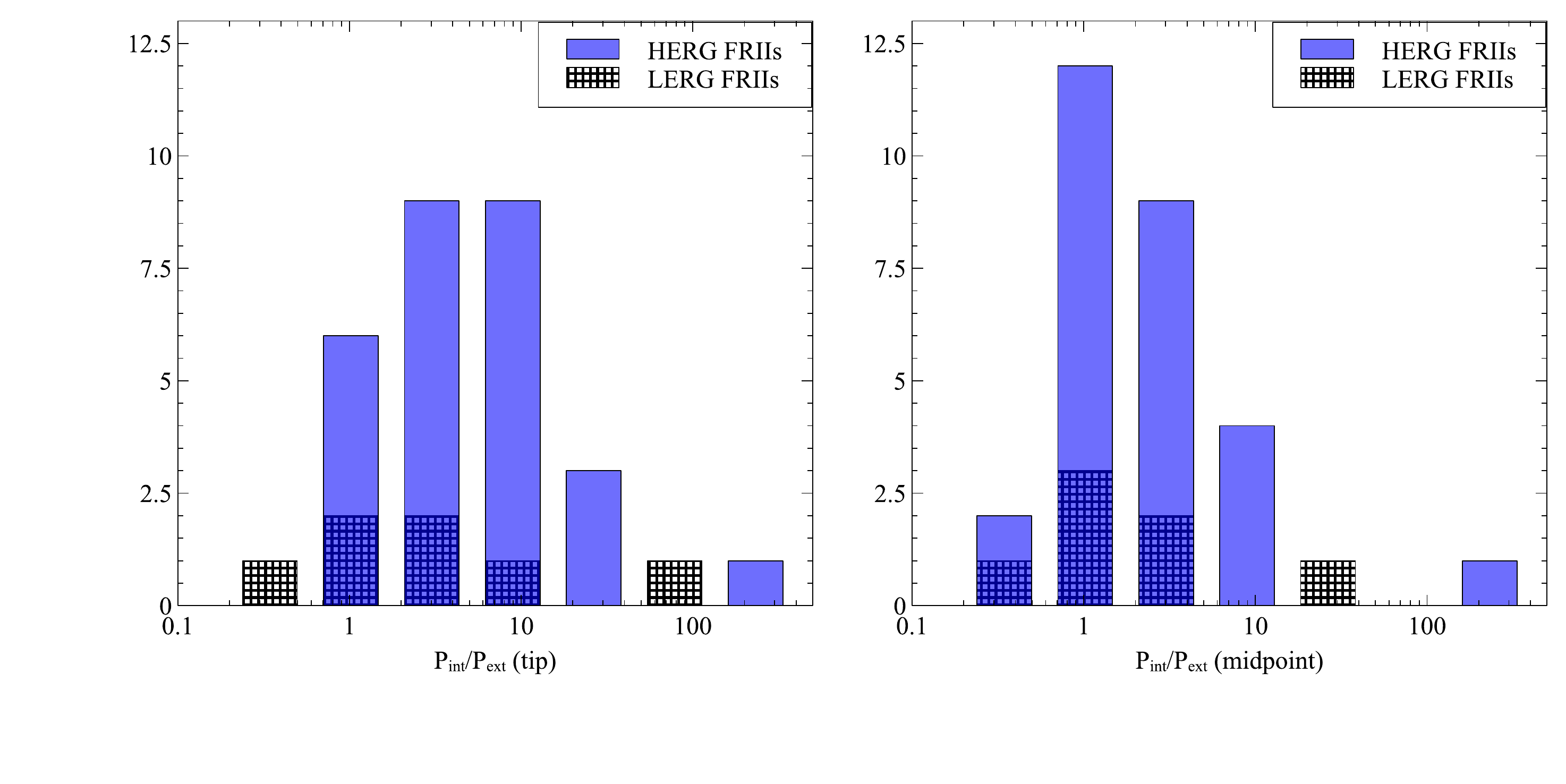}
\caption{Pressure ratio comparison for FRIIs, broken down by
    accretion mode. Low-excitation FRII radio galaxies (LERGs) are
    indicated by hatched black regions, while high-excitation FRIIs (HERGs) are
    shown in blue. Pressures are the inverse-Compton derived pressures from \citetalias{ineson17}, with pressure ratio limits omitted.}
\label{fig:accmodefrii}
\end{figure*}

The \citetalias{ineson15} LERG sub-sample contains both FRIs and FRIIs, while the HERG sample is mostly FRIIs, with only two FRIs. Therefore the simplest way to check for a LERG/HERG pressure ratio difference is to consider only the FRII subsample. Fig.~\ref{fig:accmodefrii} shows the FRII pressure ratios broken down by accretion mode. The pressure ratio distributions are indistinguishable, with nearly all of the LERG FRIIs having ratios greater than 1. Therefore there is no indication that the LERG FRIIs require a significant proton contribution, unlike the FRI subsample. A Wilcoxon-Mann-Whitney test confirms that the LERG and HERG subsamples are consistent with having been drawn from the sample parent population, which provides strong evidence that the FRI and FRII AGN sub-populations are physically different systems with different particle content -- there is no evidence that accretion mode is relevant to radio-lobe particle content on scales of tens to hundreds of kpc.

\section{Summary}

We have presented a new, systematic comparison of FRI and FRII
radio-galaxy particle content, as inferred from comparisons of
internal plasma conditions with the external pressure of the cluster
environment. We conclude that:
\begin{enumerate}
\item The distribution of radio-lobe internal energy between radiating particles, non-radiating particles and
magnetic field must be substantially different for the FRI and FRII sub-populations, with
the simplest explanation for the difference being that FRI radio
galaxies typically have higher levels of proton content. 
\item As discussed in many previous works, there is
substantial variation in the particle content of {\it both} classes of
object: some FRI radio galaxies are relatively close to pressure
balance without the need for non-radiating particles, and a small
subset of FRIIs in the richest environments appear to need
substantial non-radiating particle content. We argue that the most likely explanation is that the interplay between jet and environment leads to varying levels of proton contamination via entrainment and/or buoyant
mixing.
\item Morphology is an important source of scatter in the relationship between radio luminosity and cluster environmental richness; however, it is unclear that this can be a consequence mainly of differing particle content. It may instead be driven by differences in lobe evolution in richer environments. 
\item The relationship between jet power and radio luminosity is systematically different for the FRII jet population from that of the FRI X-ray cavity sources: cavity jet-power scaling relations may overestimate FRII jet powers by up to an order of magnitude.
\item Radio-lobe particle content is unconnected to accretion mode: the internal conditions of FRII LERG and HERG radio galaxy lobes are indistinguishable.
\end{enumerate}

Our results highlight the physical diversity of the radio-loud AGN population. It will be crucial to account for this diversity, and for the underlying demographics resulting from particular sample selection methods, in any attempts to infer physical properties for AGN populations identified in ongoing and future deep, wide-field radio surveys (e.g. the LOFAR Two-Metre Sky Survey (LOTSS): \citealt{shimwell17}, surveys with MeerKAT, ASKAP and the Square Kilometre Array). If we want to make robust estimates of the cosmological contribution of AGN feedback from jets at all redshifts, then it is essential that we use these surveys, together with simulations and detailed follow-up, to improve our understanding of the physics of radio jet evolution and the mapping between jet populations and large-scale environment.

\section*{Acknowledgments}

JHC and MJH acknowledge support from the Science and Technology Facilities Council (STFC) under grants ST/M001326/1, ST/R00109X/1, and ST/M001008/1. The plots in this paper were made using the Veusz software, written by Jeremy Sanders (\url{https://veusz.github.io/}). We would like to thank the anonymous referee for helpful comments that improved the paper.

\bibliographystyle{mnras}
\bibliography{qjet}

\label{lastpage}
\end{document}